\documentclass[prx,twocolumn,amsmath,amssymb,superscriptaddress,longbibliography]{revtex4-2}

\usepackage{color,times}
\usepackage{bm}
\usepackage{dcolumn}
\usepackage{multirow}
\usepackage{amssymb,amsfonts,amsmath,graphicx}
\usepackage{epsfig}
\usepackage{xcolor}
\usepackage{SIunits}
\usepackage{braket}
\usepackage{epstopdf}
\usepackage{physics}
\usepackage{stmaryrd}
\usepackage{bbold}
\usepackage{amsmath}
\usepackage[utf8]{inputenc}
\usepackage{chngcntr}
\usepackage[english]{babel}
\usepackage[colorlinks=true, breaklinks=true, linkcolor=darkblue, citecolor=darkblue, urlcolor=darkblue]{hyperref}
\usepackage{textcomp}
\usepackage{soul}
\usepackage{makecell}
\usepackage{todonotes}
\usepackage{multirow}
\usepackage{booktabs}   
\usepackage{array}
\usepackage{float} 

\definecolor{darkblue}{rgb}{0, 0, 0.8}
\definecolor{darkgreen}{rgb}{0, 0.5, 0}

\graphicspath{{//}}

\counterwithout{equation}{section}
\addtocounter{equation}{0}

\begin{document}
	
\title{Observation of electron spin interactions between Rydberg atoms}

\author{Le Ruan}
\thanks{These authors contributed equally to this work.}
\affiliation{Institute of Physics, Chinese Academy of Sciences, Beijing 100190, China}

\author{Ziqi Zhou}
\thanks{These authors contributed equally to this work.}
\affiliation{Institute of Physics, Chinese Academy of Sciences, Beijing 100190, China}

\author{Yuchen Guo}
\affiliation{Institute of Physics, Chinese Academy of Sciences, Beijing 100190, China}

\author{Chengshu Li}
\affiliation{Institute for Advanced Study, Tsinghua University, Beijing, 100084, China}
\affiliation{Beijing Key Laboratory of Cold Atom Quantum Computation, Tsinghua University, 100084, Beijing, China}

\author{Cheng Chen}
\email{cheng.chen@iphy.ac.cn}
\affiliation{Institute of Physics, Chinese Academy of Sciences, Beijing 100190, China}

\begin{abstract}
We report the observation of electron spin interactions between Rydberg atoms, which are driven by spin-orbit coupling through second-order dipole perturbation and exhibit a spatial anisotropy governed by the atomic configuration.
Specifically, we observe coherent electron spin exchange dynamics, with the measured coupling strength agreeing well with both numerical calculations and theoretical models. 
Furthermore, we show that global microwave dressing enables active engineering and dynamical freezing of the spin exchange by introducing a differential AC Stark shift between the participating states.
Additionally, we achieve tunability of the interaction by applying a stronger magnetic field, which effectively modifies the energy contributions of the underlying spin-orbit coupling channels. 
Finally, measuring spin dynamics in one-dimensional multi-atom chains aligned parallel or perpendicular to the magnetic field provides a self-consistent validation of the anisotropic XXZ framework.
This electron spin interaction natively features spin-position coupling and enables a natural mapping onto the Heisenberg-Kitaev model~\cite{Li2026}.
Our findings reveal a new class of electron spin-spin interactions among Rydberg atoms, expanding the scope of quantum simulation with Rydberg atom arrays.
\end{abstract}
\date{\today}

\maketitle

\begin{figure}[!htbp]
	\centering
	\includegraphics[width=0.9\linewidth]{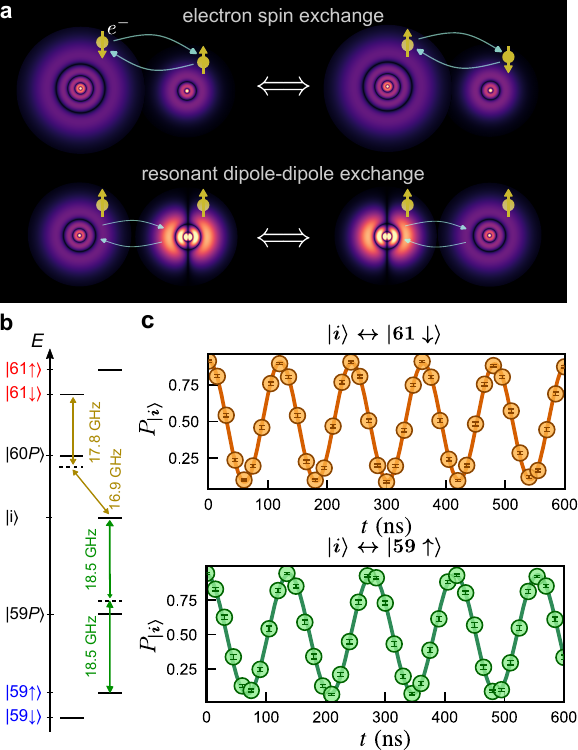}
	\caption{\textbf{Schematic diagram of the electron spin exchange, relevant Rydberg levels, and calibration of two-photon microwave $\pi$-pulse by Rabi oscillations.}
	\textbf{a}.~Electron spin exchange between Rydberg atoms with a principal quantum number difference of $\Delta n=2$, and the conventional resonant dipole-dipole exchange.
	\textbf{b}.~Relevant Rydberg levels, and two-photon microwave transitions used to prepare the initial state $\ket{61\downarrow, 59\uparrow}$.
	\textbf{c}.~Rabi oscillations of the two-photon microwave transitions.}
	\label{fig:fig_1}
\end{figure}

Quantum magnetism arises from the interplay between the spin and the spatial degrees of freedom of electrons, and in solids spin-orbit coupling can generate bond-dependent, highly anisotropic spin models such as the Kitaev honeycomb Hamiltonian~\cite{Jackeli2009,Chaloupka2010}.
Rydberg atom arrays have become a dynamic platform for quantum simulation and quantum computation~\cite{Saffman2010,Browaeys2020,Wu2021,Bluvstein2026}, combining defect-free, reconfigurable geometries~\cite{Kim2016,Barredo2016,Endres2016,Lin2025} with strong, tunable interactions and single-atom readout.
These capabilities have enabled the realization of a broad family of spin models---including the transverse-field Ising~\cite{Labuhn2016,Bernien2017,Scholl2021}, dipolar XY~\cite{Leseleuc2019, Chen2023,Bornet2023,Bornet2024,Emperauger2025,Chen2025,Emperauger2025a,Zhang2025,Yue2026}, Heisenberg XXZ~\cite{Scholl2022,Geier2021,Kim2024,Zeiher2017}, and bosonic $t$--$J$~\cite{Homeier2024,Qiao2025} models---together with applications to magnetism, entanglement, spin squeezing, quantum gates, and optimization~\cite{Zeiher2016,Shaw2024,Peper2025,Eckner2023,Jau2016,Evered2025}.
At the heart of these models lie the resonant dipole-dipole interaction, arising from substantial transition dipole moments connecting states of opposite parity (e.g., $nS$ and $nP$ states), and its second-order perturbation.
The exchange dynamics given by this dipole interaction alters the spatial orbital state of the outer-shell electron, while leaving its spin state unchanged (Fig.~\ref{fig:fig_1}a).
However, the electron spin exchange interaction between Rydberg atoms reported here operates on the electron spin degree of freedom, rather than on spatial orbital wavefunctions as in resonant dipole-dipole interactions (Fig.~\ref{fig:fig_1}a).
Mediated by spin-orbit coupling and dipole-dipole interactions, this electron spin interaction natively features a spin-position coupling~\cite{Li2026}, rendering the exchange dynamics highly sensitive to the spatial geometry of the atomic array.
Under a magnetic field, the system effectively maps onto the anisotropic XXZ model, providing a highly controllable platform to explore non-equilibrium physics, magnon bound states, and spin squeezing~\cite{Geier2021,Scholl2022,Franz2024,Kim2024,Lu2026,Block2024}.

\begin{figure*}
	\centering
	\includegraphics[width=0.9\textwidth]{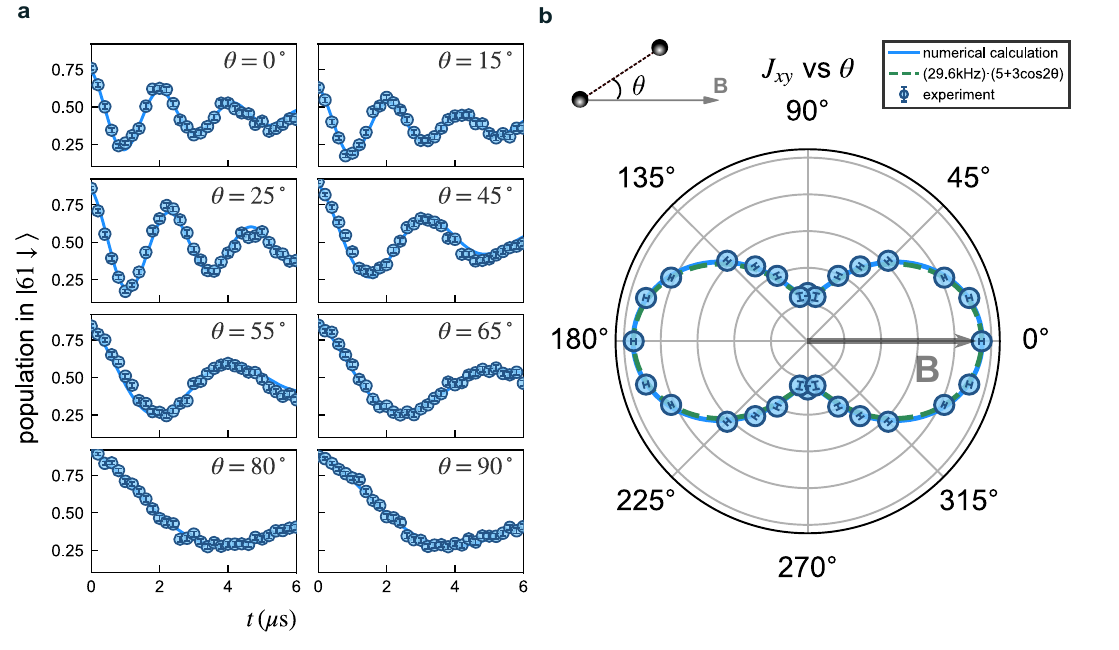}
	\caption{\textbf{Observing electron spin exchange between Rydberg atoms.}
	\textbf{a}.~Experimental data of electron spin exchange at fixed atomic separation $d=9.66~\mu\text{m}$ for various $\theta$. Solid curves represent fits to the experimental data using $P(t)=A\,e^{-t/\tau}\cos(2J_{xy}t/\hbar+\phi_0)+C$, with the same global phase offset $\phi_0$ shared across all angles. We present ideal-case models, simulations accounting for experimental imperfections and the measured data in the Supplementary Material~\cite{SM}.
	\textbf{b}.~Angular dependence of $J_{xy}$. We compare the experimental data with numerical calculation~\cite{Weber2017,Moegerle2026} (solid line) and theoretical model $J_{xy}(\theta)=J(5+3\cos 2\theta)/8$ (dashed line). We obtain $J_{xy}/h=0.236\pm 0.002~\text{MHz}$ at $\theta=0^\circ$, yielding $J_{xy}(\theta)/h = (29.6\pm 0.3)~(5+3\cos 2\theta)~\text{kHz}\,$. Error bars denote $2\sigma$ confidence intervals from the covariance of the fits. For comparison, we present in the Supplementary Material the principal-quantum-number exchange dynamics for the $\Delta n = 1$ case, where the interaction exhibits no pronounced spin-position coupling due to the absence of pivotal spin-orbit coupling effects~\cite{SM}.} 
	\label{fig:fig_2}
\end{figure*}

Here, we report the observation of electron spin exchange between Rydberg atoms, focusing on three aspects: 
First, by selecting specific Rydberg states, we observe coherent electron spin exchange dynamics between two atoms, with the extracted coupling strengths showing quantitative agreement with both numerical calculations and theoretical models.
Second, we demonstrate coherent control over the spin interaction via a differential AC Stark shift induced by microwave dressing. In addition, applying a stronger magnetic field enables tuning of the interaction by effectively modifying the energy contributions of the underlying spin-orbit coupling channels.
Third, we extend our observation to spin dynamics in one-dimensional atomic chains, validating the physical framework through a comparative study of anisotropic XXZ dynamics for configurations parallel and perpendicular to the magnetic field.
Our findings reveal that this electron spin exchange interaction is driven by spin-orbit coupling via second-order dipole perturbation, and exhibits a pronounced spatial anisotropy governed by the atomic configuration.

Our experiment is implemented in a two-dimensional $^{87}\text{Rb}$ atom array, where the spin qubits are encoded in the $m_J = \pm 1/2$ Zeeman sublevels of the $59S_{1/2}$ and $61S_{1/2}$ Rydberg manifolds for distinct atoms, labeled as $\ket{n\uparrow} = \ket{nS_{1/2, +1/2}}$ and $\ket{n\downarrow} = \ket{nS_{1/2, -1/2}}$ (Fig.~\ref{fig:fig_1}b).
Under this configuration, the resulting spin-spin coupling takes an anisotropic dipolar form~\cite{Li2026}, $ J\left[ \mathbf{S}_1 \cdot \mathbf{S}_2 - \frac{3}{2} (\mathbf{S}_1 \cdot \mathbf{n})(\mathbf{S}_2 \cdot \mathbf{n}) \right]$, with $\mathbf{n}=(\sin \theta \cos \psi, \sin \theta \sin \psi, \cos \theta)$ representing the interatomic unit vector defined by the polar angle $\theta$ and azimuthal angle $\psi$ relative to the quantization axis.
This form follows from a symmetry analysis: the dipole-dipole interaction $V_{dd}\propto R^{-3}\left[\mathbf{r}_1\cdot\mathbf{r}_2-3(\mathbf{r}_1\cdot\mathbf{n})(\mathbf{r}_2\cdot\mathbf{n})\right]$ couples the two electron coordinates $\mathbf{r}_1,\mathbf{r}_2$ to the interatomic unit vector $\mathbf{n}$, while spin-orbit coupling couples each electron spin $\mathbf{S}$ to its coordinate $\mathbf{r}$ through $\mathbf{S}\cdot(\mathbf{r}\times\mathbf{p})$; chaining these two couplings together yields a spin-spin interaction with a similar dipolar form.
In the experiment, we set a magnetic field of $21.7~\text{G}$ to lift the degeneracy of the spin sublevels, enabling high-fidelity preparation of initial state $\ket{61\downarrow, 59\uparrow}$ via global microwave and site-resolved addressing laser beams.
In this case, the magnetic field enforces an effective $U(1)$ symmetry, confining the system dynamics to the total magnetization $M = 0$ sector. 
The resulting Hamiltonian takes the XXZ form $J_{xy}\left(S_1^x S_2^x + S_1^y S_2^y\right) + J_z S_1^z S_2^z$, where the coupling terms are analytically given by $J_{xy} = J (5 + 3\cos 2\theta)/8$ and $J_z = J (1 - 3\cos 2\theta)/4$~\cite{Li2026}. 
Consequently, tuning the orientation angle $\theta$ provides direct control over the anisotropy $J_z / J_{xy}$ of the effective spin model.

We first configure atom pairs with a constant interatomic spacing of $d = 9.66~\mu\text{m}$, where the angle $\theta$ spans from $0^\circ$ to $90^\circ$ (Fig.~\ref{fig:fig_2}).
Initial state preparation of the staggered state $\ket{61\downarrow, 59\uparrow}$ relies on a light-shift-assisted microwave sequence (see Fig.~\ref{fig:fig_SM1} for more details~\cite{SM}).
All the atoms are first excited from the ground state $\ket{g}=\ket{5S_{1/2}, F=2, m_F=2}$ to the intermediate state $\ket{i} \equiv \ket{60S_{1/2}, m_J = +1/2}$ via STIRAP, and transferred to $\ket{59\uparrow}$ by a two-photon microwave $\pi$-pulse (Fig.~\ref{fig:fig_1}c). 
A $1013\text{-nm}$ addressing beam induces a local AC Stark shift $\delta$ for state $\ket{i}$ on one atom, decoupling it from the following microwave pulses.
This allows another atom to be selectively driven through $\ket{i}$ into $\ket{61\downarrow}$, while the addressing light shift remains active to suppress interaction-induced dynamics involving $\ket{i}$.
Once the staggered state $\ket{61\downarrow, 59\uparrow}$ is prepared, spin-spin interactions drive coherent electron spin exchange via the channel $\ket{61\downarrow, 59\uparrow} \leftrightarrow \ket{61\uparrow, 59\downarrow}$. 
For readout, the addressing beam is applied again, while a two-photon microwave $\pi$-pulse selectively maps state $\ket{61\downarrow}$ of the non-addressed atom back to $\ket{i}$. 
State $\ket{i}$ is optically deexcited to the ground state, which is then recaptured by the tweezers while expelling remaining Rydberg states.

Fig.~\ref{fig:fig_2}a shows the population dynamics of the state $\ket{61\downarrow}$ of the non-addressed atoms across a range of orientation angles $\theta$, directly evidencing electron spin exchange.
Fitting these coherent oscillations yields the exchange interaction energy as a function of $\theta$.
As highlighted in Fig.~\ref{fig:fig_2}b, the extracted angular dependence exhibits remarkable agreement with both exact numerical calculations~\cite{Weber2017,Moegerle2026} and the theoretical model.

\begin{figure}
	\centering
	\includegraphics[width=1\linewidth]{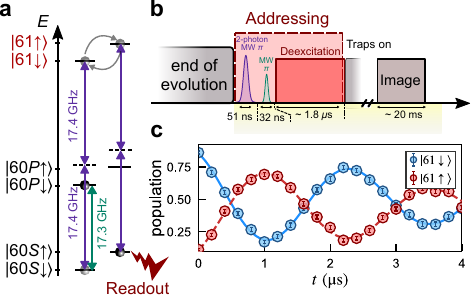}
	\caption{\textbf{Checking the population in $\ket{61\uparrow}$.}
	\textbf{a}.~Microwave transitions used for the readout of population in $\ket{61\uparrow}$, including a two-photon $\pi$-pulse transferring $\ket{61\uparrow}$ to $\ket{60\uparrow}$ and a $\pi$-pulse transferring $\ket{60\downarrow}$ to $\ket{60P\downarrow}$.
	\textbf{b}.~Readout sequence for $\ket{61\uparrow}$. Since the two-photon $\pi$-pulse can also transfer $\ket{61\downarrow}$ to $\ket{60\downarrow}$, which would also be deexcited and detected, we apply a second $\pi$-pulse driving $\ket{60\downarrow}$ to $\ket{60P\downarrow}$ to eliminate this crosstalk.
	\textbf{c}.~Experimental results for $\theta = 25^\circ$ with $\ket{60\uparrow}$ readout scheme.}
	\label{fig:fig_3}
\end{figure}

We further verify the spin-exchange flip of the non-addressed atom by implementing an alternative readout protocol targeting the population of state $\ket{61\uparrow}$ (Fig.~\ref{fig:fig_3}a).
Here, evolution is terminated by a two-photon microwave $\pi$-pulse transferring $\ket{61\uparrow}$ to $\ket{i}$, followed by a targeted $\pi$-pulse driving $\ket{60\downarrow} \rightarrow \ket{60P\downarrow}$ to eliminate crosstalk from unwanted $\ket{60\downarrow}$ detection channels (Fig.~\ref{fig:fig_3} b), where the state $\ket{60P\uparrow}$ denotes $\ket{60P_{3/2, 1/2}}$ and $\ket{60P\downarrow}$ denotes $\ket{60P_{3/2, -1/2}}$.
As shown in Fig.~\ref{fig:fig_3}c, the measured $\ket{61\uparrow}$ population dynamics mirrors the evolution of $\ket{61\downarrow}$, validating coherent $\ket{61\downarrow} \leftrightarrow \ket{61\uparrow}$ oscillations.

\begin{figure}
	\centering
	\includegraphics[width=1\linewidth]{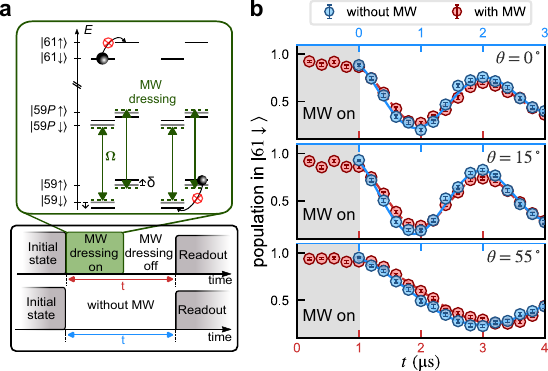}
	\caption{
	\textbf{Engineering electron spin interactions by differential AC Stark shift via microwave dressing.}
	\textbf{a}~Microwave dressing scheme: a driving field with a frequency centered between the $\ket{59\uparrow}\rightarrow\ket{59P\uparrow}$ and $\ket{59\downarrow}\rightarrow\ket{59P\downarrow}$ transitions imparts differential light shifts, thereby generating a differential energy shift between $\ket{59\uparrow}$ and $\ket{59\downarrow}$.
	\textbf{b}.~Dynamics with (shaded) and without MW dressing. The time axes of the two datasets are shifted by $1\,\mu\text{s}$ to show that spin-exchange dynamics resume immediately upon turning off the MW dressing, displaying near-perfect agreement with the baseline without MW-dressing.}
	\label{fig:fig_4_MW_dressing}
\end{figure}

\begin{figure}
	\centering
	\includegraphics[width=1\linewidth]{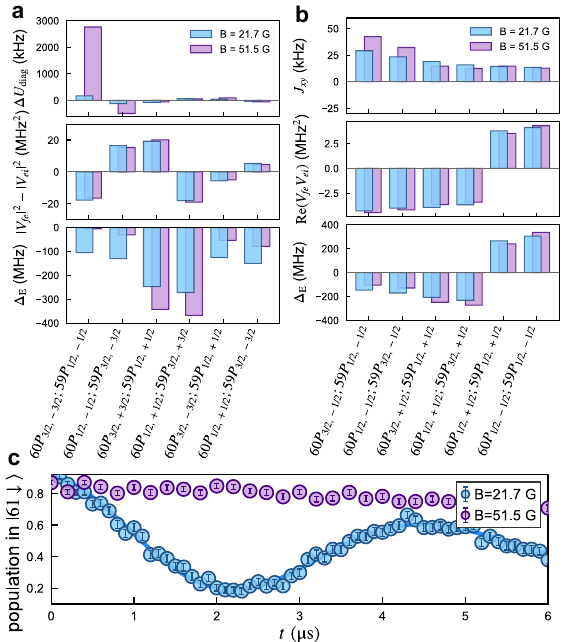}
	\caption{\textbf{Tuning electron spin interactions through contributions of perturbation channels via magnetic field.}
	\textbf{a}.~Perturbation channel-resolved decomposition of the diagonal energy splitting between the initial state $\ket{61\downarrow, 59\uparrow}$ and the final state $\ket{61\uparrow, 59\downarrow}$ for $\theta=60^\circ$ at different magnetic fields. Top panel: contributions of individual intermediate states to $\Delta U_{\text{diag}}$; middle panel: energy difference of the matrix elements coupling the initial and final states through each intermediate state, where $i$, $f$, and $e$ denote the initial, final, and intermediate states; bottom panel: detunings of the intermediate states. At $B=51.5\,\text{G}$, the detuning of this channel $\ket{60P_{3/2,-3/2};59P_{1/2,-1/2}}$ is reduced to $5.99\,\text{MHz}$ while the matrix-element difference remains essentially unchanged, causing its contribution to increase from the hundreds of kilohertz to a few megahertz level, which is sufficient to freeze the spin exchange.
	\textbf{b}.~Similar to the perturbation channel analysis in a, but for the off-diagonal coupling terms $J_{xy}$.
	\textbf{c}.~Dynamics at a magnetic field of $B=51.5\,\text{G}$ ($\theta=60^\circ$). Compared to $B = 21.7\,\text{G}$, the spin exchange is strongly suppressed, as a certain intermediate perturbation channel is shifted close to resonance.}
	\label{fig:fig_4_B_field}
\end{figure}

We next explore two distinct mechanisms for controlling the spin-exchange dynamics. 
On one hand, we utilize global microwave dressing to achieve coherent control over the interaction, where an induced differential AC Stark shift detunes the spin exchange and freezes the quantum dynamics, confirming its engineerable nature.
Specifically, at the onset of the evolution, we apply a square microwave pulse (engaged for up to $1~\mu\text{s}$) to couple the $\pi$-transitions from $59S_{1/2}$ to $59P_{1/2}$ ($\ket{59\uparrow} \rightarrow \ket{59P_{1/2}\uparrow}$ and $\ket{59\downarrow} \rightarrow \ket{59P_{1/2}\downarrow}$, where the state $\ket{59P\uparrow}$ denotes $\ket{59P_{1/2, 1/2}}$ and $\ket{59P\downarrow}$ denotes $\ket{59P_{1/2, -1/2}}$), as depicted in Fig.~\ref{fig:fig_4_MW_dressing}a. 
Set near the center of the two transition frequencies, the pulse introduces symmetric opposite detunings of approximately $\pm 20.3\,\text{MHz}$. 
With a Rabi frequency of $\Omega/2\pi = 14.8\,\text{MHz}$, it induces energy shifts on $\ket{59\uparrow}$ and $\ket{59\downarrow}$, generating a $4.9\,\text{MHz}$ differential shift that far overcomes $J_{xy}/h = 0.236\,\text{MHz}$ and suppresses population transfer (Fig.~\ref{fig:fig_4_MW_dressing}b).
During this dressing pulse, the population of state $\ket{61\downarrow}$ remains nearly frozen as shown in Fig.~\ref{fig:fig_4_MW_dressing}b. 
Once the dressing field is extinguished, the system resumes coherent spin exchange with dynamics closely matching the undressed scenario. 
The resulting time shift between the dressed (red) and bare (blue) curves directly demonstrates a microwave-induced delay of the electron spin exchange process.
On the other hand, applying a stronger magnetic field directly alters the energy denominators of the intermediate spin-orbit coupling channels. Crucially, as one specific perturbation channel is tuned close to resonance, its reduced energy denominator dramatically amplifies its contribution, driving spin dynamics that are qualitatively distinct from those observed at low field.
In Fig.~\ref{fig:fig_4_B_field}c, we present the dynamics at a stronger magnetic field of $B = 51.5\,\text{G}$ for $\theta = 60^\circ$, which exhibits a striking contrast to the behavior at $21.7\,\text{G}$. 
Under this field, the spin-exchange process is almost completely suppressed. 
As shown in Figs.~\ref{fig:fig_4_B_field}a and \ref{fig:fig_4_B_field}b, the channel-resolved decomposition illustrates how each intermediate pathway impacts the diagonal energy shift $\Delta U_{\text{diag}}$, the spin-exchange coupling $J_{xy}$, the interaction matrix elements ($\vert{}V_{fe}\vert{}^2 - \vert{}V_{ei}\vert{}^2$ and $\text{Re}(V_{fe}V_{ei})$), and the corresponding energy denominators $\Delta E$.
This suppression arises because a certain intermediate perturbation channel $\ket{61S_{1/2,-1/2};59S_{1/2,+1/2}} \rightarrow \ket{60P_{3/2,-3/2};59P_{1/2,-1/2}}$ is shifted close to resonance, rendering the simple second-order perturbative picture invalid.
Numerical calculations performed with the \textit{Pairinteraction} software package~\cite{Moegerle2026} validate our understanding: at this high magnetic field, the energy splitting between the diagonal terms $\Delta U_{\text{diag}}$ becomes predominant (Fig.~\ref{fig:fig_4_B_field}a), thereby strongly detuning and suppressing the off-diagonal spin exchange.
Importantly, we note that this electron spin exchange interaction $J_{xy}$ originates from the key role of spin-orbit coupling within the second-order dipole-dipole perturbation.
Specifically, for the chosen $\ket{61S}$ and $\ket{59S}$ manifolds, the spin-exchange coupling is dominated by second-order virtual transitions through four sets of near-resonant transition channels associated with the following fine-structure manifolds: $\ket{60P_{1/2}, 59P_{1/2}}$, $\ket{60P_{3/2}, 59P_{1/2}}$, $\ket{60P_{1/2}, 59P_{3/2}}$, $\ket{60P_{3/2}, 59P_{3/2}}$.
Their relatively small energy defect renders larger perturbative contributions compared to other energy-remote intermediate states.
More precisely, to ensure constructive interference among channels with sign-varying matrix element products $\text{Re}(V_{fe}V_{ei})$, we adopt the $\Delta n = 2$ configuration where spin-orbit coupling naturally aligns the signs of the corresponding energy denominators. 
Consequently, as shown in Fig.~\ref{fig:fig_4_B_field}b, all pathways—featuring concurrently negative values on the left four channels and positive values on the right two channels—contribute additively to the spin-exchange interaction $J_{xy}$ rather than interfering destructively~\cite{SM} (see Fig.~\ref{fig:FigSM_offdiag_comparision_delta_n} for details).
%

\begin{figure}
	\centering
	\includegraphics[width=1\linewidth]{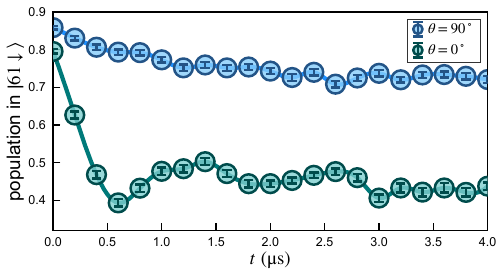}
	\caption{\textbf{Many-body dynamics in one-dimensional chains.}
	~For $\theta=0^\circ$, $J_{xy}=J$, $J_z=-J/2$, the spin polarization of non-addressed atoms relaxes rapidly into the $xy$-plane. For $\theta=90^\circ$, $J_{xy}=J/4$, $J_z=J$, the strong $z$-polarization is largely retained with less relaxation, evidencing the dominance of the longitudinal coupling $J_z$. Both chains consist of $18$ atoms at $d=9.66\,\mu\text{m}$, initialized in staggered N\'{e}el state along $z$.}
	\label{fig:fig_5}
\end{figure}

Finally, we scale the system size from two to 18 atoms arranged in a one-dimensional chain, to validate our model in a many-body system.
With the system initialized in a $z$-polarized N\'{e}el state, we track the average population of state $\ket{61\downarrow}$ for the non-addressed atoms under two orthogonal magnetic field orientations, $\theta = 0^\circ$ and $90^\circ$. 
As shown in Fig.~\ref{fig:fig_5}, at $\theta = 0^\circ$, the spin polarization rapidly relaxes from the $z$-axis into the $xy$-plane, reflecting a regime dominated by the transverse exchange coupling $J_{xy}$. 
In contrast, at $\theta = 90^\circ$, the strong $z$-polarization is largely retained with less relaxation, evidencing the dominance of the longitudinal coupling $J_z$. 
These contrasting relaxation dynamics fully align with the many-body analytical XXZ Hamiltonian, $H = \sum_{i < j} \frac{1}{R_{ij}^6} \left[ J_{xy}(\theta_{ij}) (S_i^x S_j^x + S_i^y S_j^y) + J_z(\theta_{ij}) S_i^z S_j^z \right]$, where $J_{xy}(\theta) = J (5 + 3\cos 2\theta)/8$ reaches its maximum $J_{xy} = J$ at $\theta = 0^\circ$ with $J_z = -J/2$, whereas $J_z (\theta) = J (1 - 3\cos 2\theta)/4$ dominates at $\theta = 90^\circ$ with $J_{xy} = J/4$ and $J_z = J$.

In conclusion, we have identified and realized position-dependent electron spin interactions in Rydberg atom arrays, unveiling a paradigm governed by the interplay of dipole-dipole interactions and spin-orbit coupling. 
Unlike the $\Delta n = 1$ exchange between identical $m_J = +1/2$ Rydberg $S$ states, which only swaps the principal quantum numbers~\cite{Emperauger2025a, Qiao2025}, it is the genuine electron spin that is exchanged in our scheme.
Whereas $\Delta n = 2$ Rydberg S-state pairs have previously been studied either to propose spin-exchange entanglement theoretically~\cite{Shi2014} or, near Förster resonance, to boost single-photon nonlinearities~\cite{ParisMandoki2016,Gorniaczyk2016}, the spin-orbit-coupling origin of the exchange and the resulting spin-position coupling were not identified; here we directly observe exactly this mechanism.
Our results validate the proposed spin-position-coupled interaction mechanism, as the measured time-resolved dynamics show close agreement with both the theoretical framework~\cite{Li2026} and numerical simulations from the \textit{Pairinteraction} package~\cite{Moegerle2026}.
Active control over the interaction was demonstrated by dynamically applying a differential AC Stark shift via microwave dressing, and altering contributions of the perturbation channels via an elevated magnetic field. 
Furthermore, extension to 18-atom chains benchmarked the anisotropic XXZ dynamics in the many-body regime. 
Looking ahead, distinct from the digital simulation of the Kitaev model~\cite{Evered2025}, this electron spin interaction opens promising avenues for natively realizing the Heisenberg--Kitaev model in bilayer arrays with trirectangular tetrahedral primitive cells near zero magnetic field~\cite{Li2026}, as well as accessing anisotropic Heisenberg models under high magnetic field.

{\it Note Added.}
Recently we became aware of related work based on a dual-species platform via energy spectroscopy (Peng Xu, private communication).

\begin{acknowledgments}
{\it Acknowledgments.}
We acknowledge Hui Zhai for fruitful discussions. 
We thank Qianjiao He, Tong Wu, Tianle Gu and Yuzhe Wang for early experimental contributions.
This work is supported by Quantum Science and Technology-National Science and Technology Major Project (2024ZD0301700), and the start-up grant from IOP-CAS.
\end{acknowledgments}

\bibliography{references_electron_spin_exchange}

\clearpage

\setcounter{figure}{0}
\renewcommand\thefigure{S\arabic{figure}} 
\setcounter{section}{0}
\renewcommand\thesection{S\arabic{section}}
  
\begin{center}
{\bf Supplemental Material}
\end{center}


\section{Experimental methods}\label{SM:Exp_details}

\subsection{Experimental sequence}\label{SubSM:experimental_sequence}

\begin{figure*}
	\centering
	\includegraphics[width=1\linewidth]{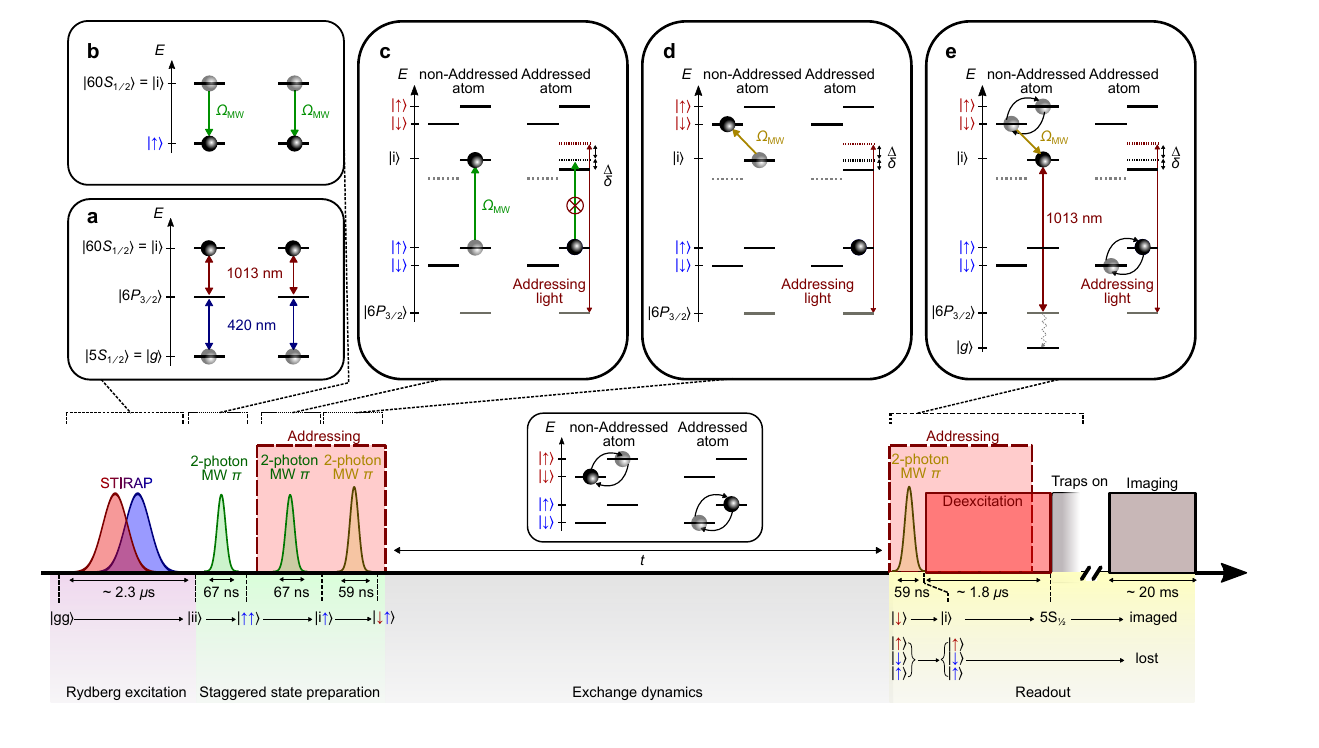}
	\caption{\textbf{Detailed experimental sequence.}
	\textbf{a}.~Two-photon stimulated Raman adiabatic passage (STIRAP) transfers the atoms from the ground state to the initial state $\ket{i}$.
	\textbf{b}.~A global microwave pulse flips both atoms from $\ket{i}$ to $\ket{59\uparrow}$.
	\textbf{c}.~A combination of microwave pulse and addressing light selectively prepares the $\ket{i,59\uparrow}$ state.
	\textbf{d}.~The atoms flipped back to $\ket{i}$ on non-addressed sites are transferred to $\ket{61\downarrow}$, completing the preparation of the staggered state $\ket{61\downarrow, 59\uparrow}$.
	\textbf{e}.~Readout pulse sequence: The addressing light is re-applied while a two-photon microwave $\pi$-pulse selectively transfers atoms in $\ket{61\downarrow}$ back to $\ket{i}$, and then to ground state for final imaging.
		}
	\label{fig:fig_SM1}
\end{figure*}

Figure~\ref{fig:fig_SM1} shows the complete experimental sequence from Rydberg excitation to readout. 
After rearrangement of the atom array, Raman sideband cooling to $\sim 10\,\mu$K, and optical pumping to $\ket{g}=\ket{5S_{1/2}, F=2, m_F=2}$, we adiabatically ramp down the trap depth by a factor of $\sim 50$ further cooling to $ 2.4\,\mu$K and switch off the trapping light. 
Using two-photon stimulated Raman adiabatic passage (STIRAP) with 420-nm and 1013-nm lasers via the intermediate state $\ket{6P_{3/2}, F=3, m_F=3}$, we transfer all atoms from the ground state to the target Rydberg state $\ket{60S_{1/2}, m_J=+1/2}$ (denoted as $\ket{i}$), with a total pulse duration of approximately $2.3\,\mu$s (as shown in Fig.~\ref{fig:fig_SM1}b).
Following Rydberg excitation, we prepare the initial state through a sequence of microwave pulses combined with site-resolved addressing light shifts. 
First, we apply a two-photon microwave $\pi$-pulse detuned by approximately $100\,$MHz from the $\ket{59P}$ level (duration $67\,$ns), transferring all atoms from $\ket{i}=\ket{60S_{1/2}, m_J=+1/2}$ to $\ket{59S_{1/2}, m_J=+1/2}$ (denoted as $\ket{59\uparrow}$) (as shown in Fig.~\ref{fig:fig_SM1}c). 
We then turn on the addressing light, which is 1013-nm light detuned from the $\ket{6P_{3/2}}\leftrightarrow\ket{60S_{1/2}}$ transition by $\Delta/(2\pi)\sim 420\,$MHz, resulting in a light shift $\delta\sim \Omega_{1013}^2/(4\Delta)$ on the addressed atom in the $\ket{60S_{1/2}}$ state, where $\Omega_{1013}$ is the Rabi frequency. 
At this point, the two-photon microwave $\pi$-pulse is no longer resonant with the addressed atoms, which remain in $\ket{59\uparrow}$; the non-addressed (free) atoms, however, undergo a resonant two-photon microwave $\pi$-pulse (duration $67\,$ns) back to $\ket{i}$ (as shown in Fig.~\ref{fig:fig_SM1}d). 
Immediately after, we apply another two-photon microwave $\pi$-pulse detuned by approximately $100\,$MHz from the $\ket{60P}$ level (duration $59\,$ns), further transferring the non-addressed atoms to $\ket{61S_{1/2}, m_J=-1/2}$ (denoted as $\ket{61\downarrow}$, as shown in Fig.~\ref{fig:fig_SM1}e). 
It is important to note that after the transfer to $\ket{59\uparrow}$, the addressing light must remain on until the initial state preparation is complete—this is because when the non-addressed atoms return to $\ket{i}$, interactions between addressed and non-addressed atoms would otherwise occur, and the light shift from the addressing beam is required to suppress these interactions. 
At this point, the staggered initial state is prepared: non-addressed atoms are in $\ket{61\downarrow}$, and addressed atoms are in $\ket{59\uparrow}$.
Once the initial state is prepared, we switch off the addressing light and the system evolves under the spin-spin interactions. 
During this process, the two atoms undergo a genuine electron spin exchange: the non-addressed atom transfers from $\ket{61\downarrow}=\ket{61S_{1/2}, m_J=-1/2}$ to $\ket{61\uparrow}=\ket{61S_{1/2}, m_J=+1/2}$, while the other atom transfers from $\ket{59\uparrow}=\ket{59S_{1/2}, m_J=+1/2}$ to $\ket{59\downarrow}=\ket{59S_{1/2}, m_J=-1/2}$. 
Notably, the principal quantum number $n$ and the orbital angular momentum quantum number $l$ of both atoms remain unchanged; only the electron spin state is altered.
To avoid residual dipole-dipole interactions during the readout procedure, we introduce a decoupling step. 
At the end of the evolution, we turn the addressing light back on and apply a two-photon microwave $\pi$-pulse (duration $59\,$ns) between $\ket{61\downarrow}$ and $\ket{i}$, transferring the non-addressed atoms in $\ket{61\downarrow}$ back to $\ket{i}$. 
The addressing light remains on during this pulse to prevent interactions between the addressed atoms and the non-addressed atoms that are now in $\ket{i}$ (as shown in Fig.~\ref{fig:fig_SM1}f). 
Subsequently, we apply a $1.8\,\mu$s pulse of 1013-nm light resonant with the intermediate state $\ket{6P_{3/2}}$ to deexcite atoms from $\ket{i}$ to $\ket{6P_{3/2}}$. 
Since the lifetime of $\ket{6P_{3/2}}$ is only $113\,$ns, atoms in this state rapidly decay spontaneously to the ground state $\ket{5S_{1/2}}$. 
After the deexcitation pulse, all atoms that were in $\ket{i}$ have returned to the ground state, while atoms in other Rydberg states remain in the Rydberg manifold. 
Finally, we turn the trapping light back on: ground-state atoms are recaptured by the attractive optical dipole trap, while Rydberg-state atoms are expelled due to the ponderomotive anti-trapping potential. 
We then turn on the fluorescence imaging light and perform site-resolved fluorescence imaging with an exposure time of $20\,$ms. 
A recaptured atom corresponds to the $\ket{61S_{1/2}, m_J=-1/2}$ state, while a lost atom corresponds to the other Rydberg states.

\section{Experimental imperfections}\label{SM:Exp_imperf}

\begin{figure*}
	\centering
	\includegraphics[width=1\linewidth]{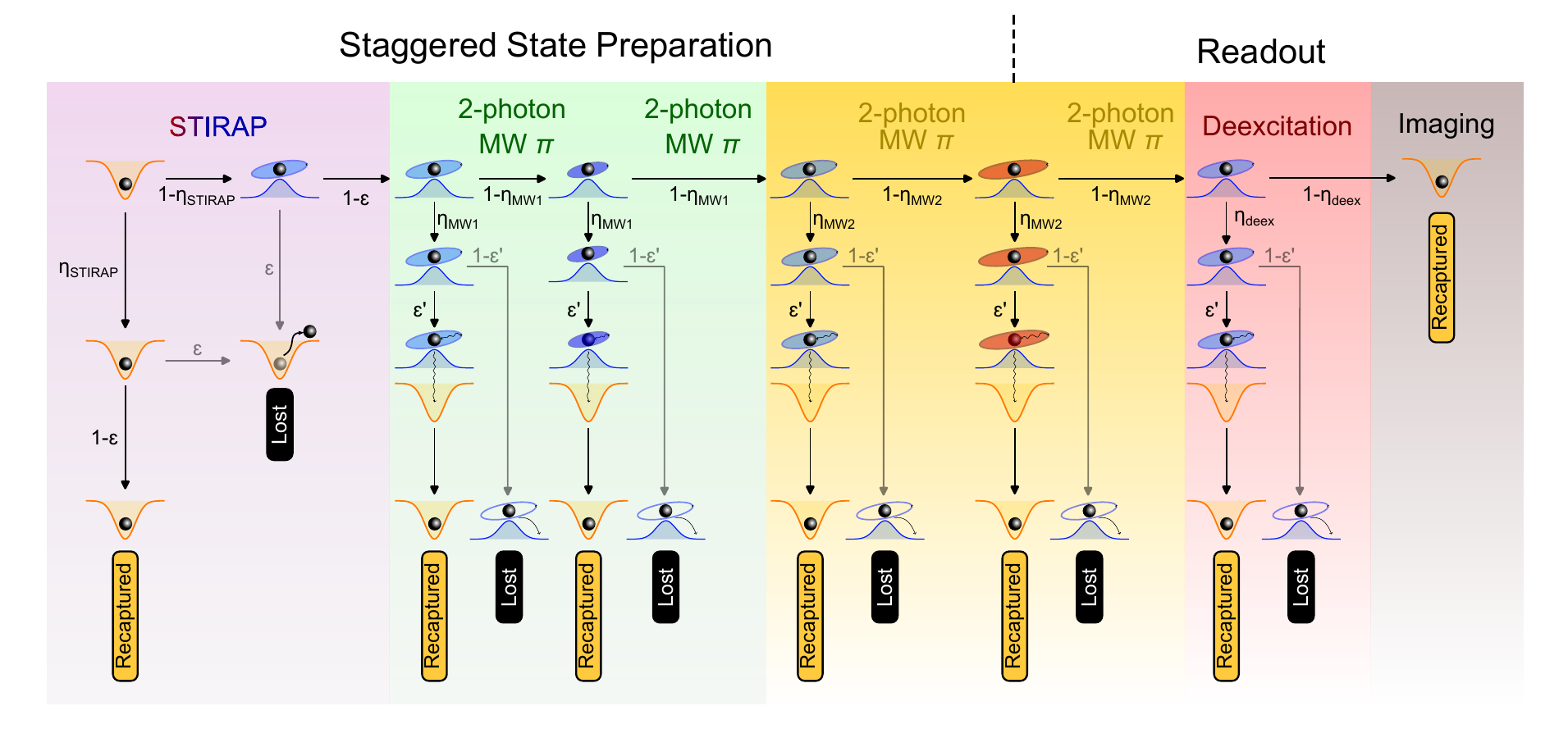}
	\caption{\textbf{Simplified error tree.}}
	\label{fig:fig_error_tree}
\end{figure*}

In this section, we review the main sources of errors that reduce the preparation and detection fidelities of the states studied in the main text. 
Due to the limitations of the readout sequence, only the population of a specific state can be measured in a single experimental run; we therefore focus our analysis on the errors associated with the target state.
Figure~\ref{fig:fig_error_tree} illustrates all the state-transfer paths experienced by the non-addressed atoms throughout the complete experimental sequence. 
To quantitatively estimate the state preparation and measurement (SPAM) errors, we decompose the sequence into a series of elementary steps, each having a finite error probability \(\eta_i\). 
Table~\ref{Tab:errors} lists the corresponding values of each \(\eta_i\), which are mainly inferred from dedicated experimental measurements, while the parameters related to the Rydberg state lifetime are estimated via numerical simulations; the last column of the table specifies the physical origins of these errors.

\begin{table*}
	\centering
	\begin{tabular}{c c c c c}
		\toprule
		Stage & Step & Symbol & Value & \multicolumn{1}{c}{Main physical origin} \\
		\midrule
		\multirow{6}{*}{Staggered state preparation} 
		&&&& Imperfect optical pumping \\
		& Rydberg excitation & $\eta_{\mathrm{STIRAP}}$ & 0.023 & Laser phase noise \\
		&&&& Spontaneous emission from $6P_{3/2}$ \\
		\cmidrule{2-5}
		& $60S\leftrightarrow59S$ MW & $\eta_{\mathrm{MW1}}$ & 0.005 & Effect of interaction during pulse \\
		\cmidrule{2-5}
		& $60S\leftrightarrow61S$ MW & $\eta_{\mathrm{MW2}}$ & 0.02  & Effect of interaction during pulse \\
		\midrule
		\multirow{5}{*}{Readout} 
		& Deexcitation & $\eta_{\mathrm{deex}}$  & 0.011 & Mechanical effect of deexcitation laser beam \\
		\cmidrule{2-5}
		& False $\ket{g}$ & $\epsilon$          & 0.022 & Atom temperature loss \\
		&&&& Background gas collisions ~\cite{Leseleuc2018} \\
		\cmidrule{2-5}
		& False $\ket{i},\ket{\uparrow},\ket{\downarrow}$ & $\epsilon'$ & 0.05  & Rydberg state radiative lifetime ~\cite{Leseleuc2018} \\
		\bottomrule
	\end{tabular}
	\caption{
	\textbf{Summary of the experimental errors defined in Fig.~\ref{fig:fig_error_tree}, together with their main physical origin.}}
	\label{Tab:errors}
\end{table*}

\subsection{State preparation errors}\label{SubSM:State_prep_imperf}

The Rydberg excitation fidelity is \((1-\eta_{\text{STIRAP}})=97.7\%\), meaning that \(2.3\%\) of the atoms remain in the ground state after the STIRAP sequence and thus do not participate in the subsequent dynamics. 
In addition, atom loss induced by finite temperature and background gas collisions amounts to \(\epsilon=2.2\%\), giving a fraction \(\eta_{\text{STIRAP}}(1-\epsilon)\) of non-interacting atoms that are eventually read out as $\ket{61\downarrow}$ at the end of the sequence.
Subsequently, each two-photon microwave \(\pi\) pulse has a finite state-transfer efficiency \((1-\eta_{\text{MW}})\), i.e. a fraction \(\eta_{\text{MW}}\) of atoms remains in their original state. Since this fraction is small, we neglect its contribution to the dynamics; these atoms are read out as $\ket{61\uparrow}$ at the end of the sequence.

\subsection{Readout imperfections}\label{SubSM:Readout_imperf}
During the readout stage, atoms that return to the $\ket{i}$ state are deexcited back to the ground state with a fidelity \((1-\eta_{\text{deex}})=98.9\%\), leaving \(1.1\%\) of the atoms still in the Rydberg state. 
Meanwhile, the finite Rydberg lifetime causes spontaneous decay to the ground state before imaging, where the atoms can be recaptured by the optical tweezers; for \(n\sim60\), this error is \(\epsilon'=5\%\).
Combining all the above errors, the initial state preparation success rate for non-addressed atoms is:
\[
(1-\eta_{\text{STIRAP}})(1-\eta_{\text{MW1}})^2(1-\eta_{\text{MW2}}) = 0.948.
\]
The total error \(\eta_{\text{non-addressed}}\) for detecting the atom in $\ket{61\downarrow}$ is given by:
\[
\begin{aligned}
\eta_{\text{non-addressed}} &= \eta_{\text{STIRAP}} \epsilon \\
&\quad + (1 - \eta_{\text{STIRAP}}) \epsilon \\
&\quad + (1 - \eta_{\text{STIRAP}})(1 - \epsilon)\eta_{\text{MW1}}(1 - \epsilon') \\
&\quad + (1 - \eta_{\text{STIRAP}})(1 - \epsilon)(1 - \eta_{\text{MW1}})\\
&\qquad\times\eta_{\text{MW1}}(1 - \epsilon') \\
&\quad + (1 - \eta_{\text{STIRAP}})(1 - \epsilon)(1 - \eta_{\text{MW1}})^2\\
&\qquad\times\eta_{\text{MW2}}(1 - \epsilon') \\
&\quad + (1 - \eta_{\text{STIRAP}})(1 - \epsilon)(1 - \eta_{\text{MW1}})^2\\
&\qquad\times(1 - \eta_{\text{MW2}})\eta_{\text{MW2}}(1 - \epsilon') \\
&\quad + (1 - \eta_{\text{STIRAP}})(1 - \epsilon)(1 - \eta_{\text{MW1}})^2\\
&\qquad\times(1 - \eta_{\text{MW2}})^2\eta_{\text{deex}}(1 - \epsilon') \\
&=7.6\%.
\end{aligned}
\]
\section{Numerical simulations}\label{SM:Simulations}

\begin{figure}
	\centering
	\includegraphics[width=1\linewidth]{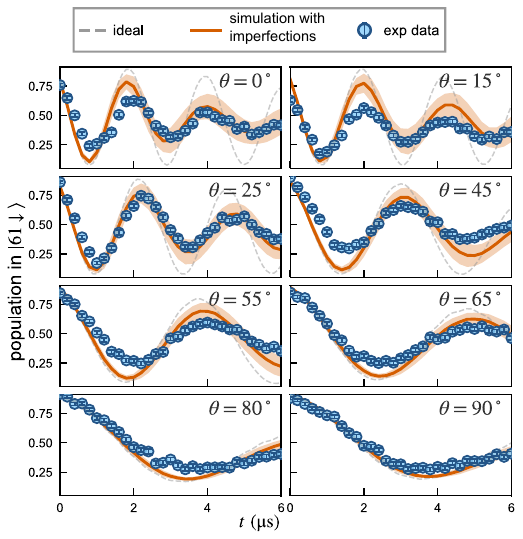}
	\caption{
	\textbf{Numerical simulation of the electron spin exchange dynamics.}
	Gray dashed curves: ideal dynamics. Red solid curves: dynamics with temperature effect and experimental imperfections (see text). Shaded areas: $2\sigma$ confidence interval ($\pm 2\times$SEM) of the Monte Carlo sampling.}
	\label{fig:fig_SM4}
\end{figure}

We perform numerical simulations of the electron spin exchange dynamics by strictly following the experimental sequence described in the previous section. 
The two-atom system, initialized in the staggered state $\ket{61\downarrow, 59\uparrow}$, evolves under the full interaction Hamiltonian; the relevant coupling strengths are computed using the \textit{Pairinteraction} software package~\cite{Moegerle2026}.
Figure~\ref{fig:fig_SM4} presents the simulated exchange oscillations. 
The gray dashed curves show the dynamics expected under ideal experimental conditions, while the red solid curves incorporate the effect of the finite atomic temperature and the experimental imperfections (SPAM errors) quantified in Sec.~\ref{SM:Exp_imperf}. 
After the adiabatic ramp-down of the optical trap, the atomic temperature is $2.4\,\mu\text{K}$, corresponding to a positional uncertainty of $\sigma_x\sim 125\,\text{nm}$ for each atom. 
In the relative coordinate of the atom pair, this translates to fluctuations of the interatomic separation $\sigma_d\sim0.18\,\mu\text{m}$ and of the orientation angle $\sigma_\theta\sim1^\circ$. 
Since the interaction strength scales as $J(\theta)\propto(5+3\cos2\theta)/r^6$, these positional fluctuations induce shot-to-shot variations of the exchange coupling with a relative standard deviation $\sigma_J/J\sim0.11$. 
A Monte Carlo sampling over these distributions yields the red curves, with the shaded areas indicating the $2\sigma$ confidence interval ($\pm 2\times$SEM) of the Monte Carlo sampling.
The simulated population contrast is further rescaled to account for the SPAM errors described in the previous section. 
From the error-tree analysis (Fig.~\ref{fig:fig_error_tree}), the total SPAM errors for a non-addressed atom is $\eta_{\text{non-addressed}}=0.076$, giving a maximum contrast $C_{\text{max}}=1-\eta_{\text{non-addressed}}=0.924$. 
This value includes the staggered-state preparation error (corresponding to a preparation fidelity of $0.948$) together with the additional readout losses, primarily deexcitation failure ($\eta_{\text{deex}}=0.011$) and atom loss from finite temperature ($\epsilon=0.022$). 
Accordingly, we set $C_{\text{min}}=\eta_{\text{non-addressed}}=0.076$, which represents the total recapture error of a non-addressed atom. 
The simulated population $P_{\text{sim}}$ is mapped to the corrected population via $P_{\text{corr}} = C_{\text{min}} + (C_{\text{max}} - C_{\text{min}})\,P_{\text{sim}}$. 
Including both the temperature-induced positional disorder and the SPAM contrast correction, the simulation reproduces the experimentally observed damping of the oscillations, indicating that positional disorder from the finite temperature is the dominant source of decoherence.

\section{Perturbation channel-resolved decomposition of the electron spin exchange interaction $J_{xy}$}\label{SM:Perturbation}
Here, we present the channel-resolved decomposition of the off-diagonal electron spin exchange interaction $J_{xy}$ across various principal quantum number configurations, providing more insights into the decisive role of spin-orbit coupling.
For each configuration of $(n_1, n_2)$, the panels display the individual channel energy contributions, the interaction matrix elements $\text{Re}(V_{fe}V_{ei})$, and the corresponding energy denominators $\Delta E$. 
In the chosen $\Delta n = 2$ configuration, the dominant near-resonant channels associated with ($59P_{1/2}$, $60P_{3/2}$) and ($59P_{3/2}$, $60P_{1/2}$) intermediate states possess positive matrix elements, while ($59P_{1/2}$, $60P_{1/2}$) possess negative matrix elements. 
Mediated by spin-orbit coupling, the energy denominators $\Delta E$ for the leading pathways (e.g., via $60P_{1/2}$) are matched in sign with $\text{Re}(V_{fe}V_{ei})$, leading to a sign alignment across all six primary channels. 
This sign matching ensures additive, constructive interference that substantially enhances the exchange interaction.
In contrast, for $\Delta n = 1$ and $\Delta n = 3$, the energy denominators remain entirely negative or positive, while the matrix elements vary in sign. 
This mismatch induces destructive interference (cancellation among channels), suppressing the net interaction.
These comparisons highlight the crucial role of spin-orbit coupling induced fine-structure splitting in engineering constructive interference and allowing electron spin exchange.

\begin{figure*}
	\centering
	\includegraphics[width=0.9\linewidth]{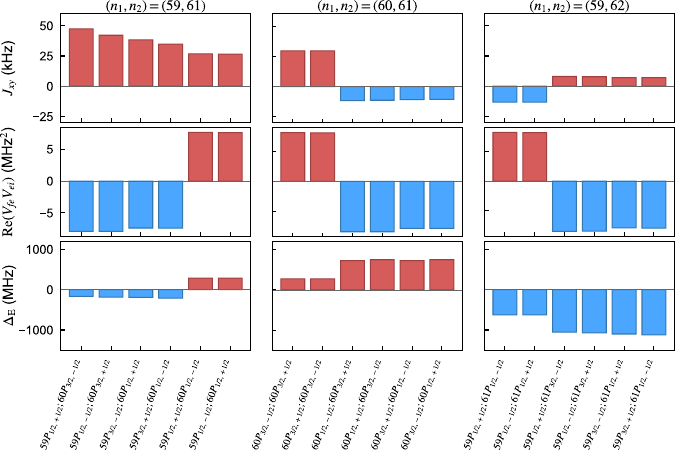}
	\caption{\textbf{Interference mechanisms of intermediate perturbation channels for different quantum-number configurations.}
	Perturbation channel-resolved decomposition of the exchange interaction $J_{xy}$ for three configurations of principal quantum numbers: $(n_1, n_2) = (59, 61)$ ($\Delta n = 2$, left), $(60, 61)$ ($\Delta n = 1$, middle), and $(59, 62)$ ($\Delta n = 3$, right). }
	\label{fig:FigSM_offdiag_comparision_delta_n}
\end{figure*}

\section{Comparison with the case of $\Delta n = 1$ }\label{SM:Comparison_delta_n}

\begin{figure}
\centering
\includegraphics[width=0.9\linewidth]{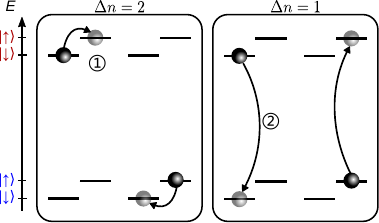}
\caption{\textbf{Comparison of exchange channels for $\Delta n = 2$ and $\Delta n = 1$.}
For $\Delta n = 2$, the initial state $\ket{61\downarrow,59\uparrow}$ is coupled to $\ket{61\uparrow,59\downarrow}$: the two atoms exchange their electron spin while each remains in its own principal-quantum-number manifold. The exchange acts on the electron spin only.
For $\Delta n = 1$, the initial state $\ket{61\downarrow,60\uparrow}$ is coupled to $\ket{60\downarrow,61\uparrow}$: with identical spin projections, the atoms instead exchange their principal quantum numbers ($60S\leftrightarrow 61S$), leaving the electron spin untouched.}
\label{fig:exchange_channels}
\end{figure}

\begin{figure*}
\centering
\includegraphics[width=0.9\linewidth]{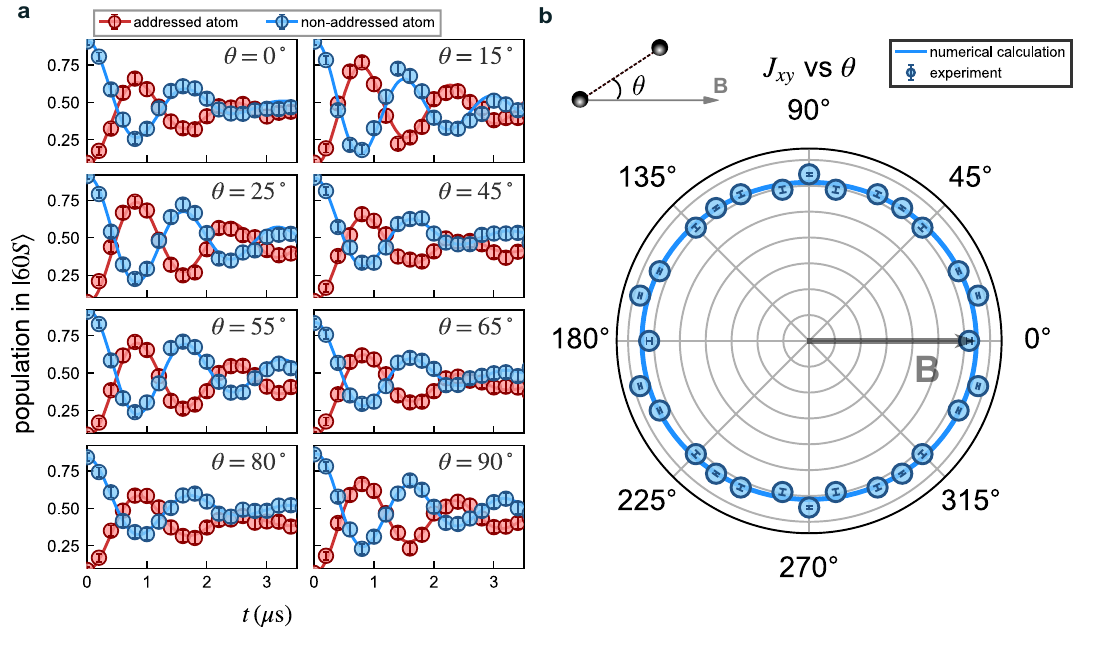}
\caption{\textbf{Angular dependence of the $\Delta n = 1$ principal-quantum-number exchange.}
	\textbf{a}.~Measured population dynamics of the $\ket{61\downarrow,60\uparrow}$ state for different orientations $\theta$ of the atom pair, showing coherent exchange to $\ket{60\downarrow,61\uparrow}$.
	\textbf{b}.~Extracted exchange coupling as a function of $\theta$. Unlike the $\Delta n = 2$ electron spin exchange, which exhibits the strong anisotropy, the $\Delta n = 1$ exchange shows no pronounced angular dependence, consistent with the absence of a decisive spin-orbit-coupling contribution.}
\label{fig:exchange_angle_dependance_60S61S}
\end{figure*}

To further clarify that the interaction reported in the main text is a genuine \emph{electron spin} exchange rather than an exchange of spatial (orbital) degrees of freedom, we compare our $\Delta n = 2$ configuration with the case of $\Delta n = 1$, in which the two atoms occupy the $60S_{1/2}$ and $61S_{1/2}$ manifolds. The distinction is most transparent at the level of the exchange process, as illustrated in Fig.~\ref{fig:exchange_channels}.
For $\Delta n = 2$, the initial staggered state $\ket{61\downarrow,59\uparrow}$ is coupled to $\ket{61\uparrow,59\downarrow}$. Here the two atoms exchange their spin projections ($m_J = -1/2 \leftrightarrow +1/2$) while each atom remains in its original principal-quantum-number manifold: atom 1 stays in $61S$, atom 2 stays in $59S$. 
The exchange therefore acts on the \emph{electron spin} degree of freedom only, with $n$ and $l$ of both atoms unchanged. 
In contrast, for $\Delta n = 1$ the initial state $\ket{61\downarrow,60\uparrow}$ is coupled to $\ket{60\downarrow,61\uparrow}$.
Since both atoms retain their own spin projection, what is exchanged here is the \emph{principal quantum number} itself: the $60S$ and $61S$ characters swap between the two atoms, while the electron spin remains a spectator.
These two cases thus correspond to the exchange of two fundamentally different degrees of freedom---the electron spin versus the principal quantum number.
This distinction is confirmed experimentally by the angular dependence of the exchange dynamics in the $\Delta n = 1$ case, shown in Fig.~\ref{fig:exchange_angle_dependance_60S61S}. For an initial state  $\ket{61\downarrow,60\uparrow}$, the population oscillates between $\ket{61\downarrow,60\uparrow}$ and  $\ket{60\downarrow,61\uparrow}$ for various orientations of the atom pair. 
In contrast to the $\Delta n = 2$ electron spin exchange, where the coupling exhibits a pronounced spin-position anisotropy $J_{xy}(\theta)=J(5+3\cos2\theta)/8$, the measured $\Delta n = 1$ dynamics show no strong angular dependence. 
Because spin-orbit coupling does not play a decisive role in the $\Delta n = 1$ channel, the exchange of the principal quantum number is nearly isotropic. 
The absence of strong angular anisotropy in the $n$-exchanging case is thus a direct experimental fingerprint that distinguishes the genuine electron spin interaction from the orbital exchange.

\end{document}